\documentstyle[12pt]{article}
\begin{document}
\title{Signature Transition and Compactification}
\author{Morteza Mohseni\thanks{\small email:mohseni@cc.sbu.ac.ir}
\\ \small Physics Department, Shahid Beheshti University, Evin ,Tehran 19839, Iran}
\date{}
\maketitle
\begin{abstract}
It is shown that a change in the signature of the space-time metric together
with compactification of internal dimensions could occur in a six-dimensional
cosmological model. We also show that this is due to interaction with Maxwell
fields having support in the internal part of the space-time.
\end{abstract}
\small{PACS:  0.420.Gz,0.4.50+h}     \small{Keywords:  signature change,higher dimensions}
\vspace{20mm}\\ \indent It is generally supposed in quantum
cosmology, either in no-boundary proposal or tunneling model that,
the universe has experienced a transition in the signature of its
metric from Euclidean to Lorentzian. It is argued that this change
of signature prevents the appearance of singularities as a result
of fluctuations in the space-time topology.

There have been recent attempts to study this effect in the
framework of classical general relativity \cite{de}. There are
also studies of the effect in the framework of higher dimensional
models \cite{ke}. In particular in \cite{em} it was suggested that
signature transition could serve as a mechanism for
compactification of the internal dimensions. This idea was pursued
further in \cite{da} in which a change in the signature of the
space-time metric induces compactification of the internal
dimension in a cosmological model with a negative cosmological
constant and no matter fields. In both of these works, it was
assumed that the internal space has a compact topology (for
example $S^1$ in \cite{da}) ab initio, but its size becomes
unobservabley small as a result of the dynamics being induced by
signature change. Now it is also interesting to ask whether
signature transition could have any role in a {\em true dynamical
compactification} \cite{ti}, that is in a scheme in which one
starts with an internal space which may or may not be compact and
compactifies it through some dynamics.

In this paper we use the results of \cite{fe} and present a model in which
both a change in the signature of the four dimensional part of space-time metric
and compactification of the internal dimensions occur as a result of interaction
of gravity with an electromagnetic field living in the internal space. We start
with a noncompact internal manifold. A change in the signature of the metric
of the internal manifold occurs and this induces a signature transition in the
four dimensional metric via interaction with the Maxwell fields leading to the
compactification of the internal dimensions. It should be added that the field
equations alone cannot induce compactification since they constrain the
manifold only locally. It is then obvious that some extra assumptions are implied
here \cite{mc}.

Here we consider a model in which Maxwell fields interact with gravity. The
action is
\begin{equation}
S = -\frac{1}{16{\pi}G}\int{R{\sqrt{-g}}d^6x} - \frac{1}{4}
\int{F_{\mu\nu}F^{\mu\nu}\sqrt{-g}d^6x},
\end{equation}
leading to the field equations
\begin{equation}
R^{\mu\nu}- \frac{1}{2}g^{\mu\nu}R=-8{\pi}GT^{\mu\nu},
\end{equation}
\begin{equation}
\frac{1}{\sqrt{-g}}\partial{\mu}(\sqrt{-g}F^{\mu\nu})=0,
\end{equation}
where $T^{\mu\nu}$ are the components of the energy momentum tensor
associated with the Maxwell fields
\begin{equation}
T^{\mu\nu} = F^{\mu}_{\hspace{1mm}\rho}F^{\nu\rho}-\frac{1}{4}F_
{\alpha\beta}F^{\alpha\beta}g^{\mu\nu}.
\end{equation}
We are interested in solutions which are of the form of the product of a
2-dimensional (semi-)Riemannian manifold $M_{2}$ and a 4-dimensional one ,
$M_{4}$ ,with the following metric
\begin{equation}
ds^2 = ds^2_2 + ds^2_4,
\end{equation}
where choosing the chart $(x^1,...,x^6)$ we may write $ds^2_2$ and $ds^4_2$ as
\begin{equation}
ds^2_2 = g_{mn}dx^mdx^n,
\end{equation}
\begin{equation}
ds^2_4 = g_{pq}dx^pdx^q,
\end{equation}
with $m,n=5,6$ and $p,q=1,2,3,4.$
In the above equations $g_{mn}$ are functions of $x^5,x^6$ while
$g_{pq}$ are functions of $x^1,x^2,x^3,x^4$ with $(x^1,...,x^4)$ and $(x^5,x^6)$
being charts on $M_4$ and $M_2$ respectively.

Maxwell equations ((3) together with $\partial_{[\lambda}F_{\mu\nu]}=0$ )
admit the following solution
\begin{equation}
F^{\mu\nu} = \frac{f{\epsilon}^{\mu\nu}}{\sqrt{|g_2|}},
\end{equation}
in which ${\epsilon}^{\mu\nu}$ is the same as ${\epsilon}^{mn}$ whenever $\mu$
and $\nu$ take the the values $m,n$ and vanishes whenever they take the values $p,q$.
Here ${\epsilon}^{mn}$ is the alternating tensor with
${\epsilon}^{56}=1$ , $g_{2}$ is the determinant of $g_{mn}$ and $f$ is a
constant.
The above solution represents electromagnetic fields living only
in the internal space. For this $F^{\mu\nu}$ we have
\begin{equation}
T^{mn} = F^m_{\hspace{2mm}\rho}F^{n\rho}- \frac{1}{4}F_{\alpha\beta}F^{\alpha\beta}g^{mn},
\end{equation}
\begin{equation}
T^{pq} = -\frac{1}{4}F_{\alpha\beta}F^{\alpha\beta}g^{pq},
\end{equation}
where $\alpha$,$\beta$$=1,...,6$. Thus the Einstein's field equations (2) become
\begin{equation}
R^{mn}- \frac{1}{2}g^{mn}R = -8{\pi}G(F^m_{\hspace{2mm}\rho}F^{n\rho}
- \frac{1}{4}F_{\alpha\beta}F^{\alpha\beta}g^{mn}),
\end{equation}
\begin{equation}
R^{pq}- \frac{1}{2}g^{pq}R= -8{\pi}G(\frac{-1}{4}F_{\alpha\beta}F^{\alpha\beta}g^{pq}).
\end{equation}
The r.h.s of equation (12) and the second term in the r.h.s of (11) are
obviously proportional to some overall cosmological terms. The first term in the
r.h.s of (11) which may be written as $F_{\sigma\rho}F^{n\rho}g^{m\sigma}$
will also be proportional to $g^{mn}$ after taking (8) into account, and hence is a
cosmological term. Thus the Maxwell fields induce cosmological constants with
different values on $M_2$ and $M_4$.

Now inserting (8) into (11) and (12) and noting that
$R_4=g_{pq}R^{pq}$ and $R_2=g_{mn}R^{mn}$ are the scalar
curvatures of $M_4$ and $M_2$ respectively, so that $R=R_2+R_4$,
we obtain the following relations
\begin{equation}
R_4={\lambda}sign(g_2),
\end{equation}
\begin{equation}
R_2=-\frac{3}{2}{\lambda}sign(g_2),
\end{equation}
in which $\lambda=8{\pi}Gf^2$ and $sign(g_2)=\frac{|g_2|}{g_2}.$
The signs of $R_4$ and $g_2$ are the same while the sign of $R_2$
is the opposite to that of $g_2$. Therefore when the time
dimension is in $R_4$ , that is $g_2>0$, the internal space $M_2$
is compact and vice versa. This is because a negative (in our
notation) scalar curvature implies that the manifold is compact
which is itself a consequence of certain theorems in global
differential geometry \cite{mc1}. Therefore equations (13) and
(14) imply spontaneous compactification of the internal
dimensions.

Now consider the following as the metric of the space-time
\begin{equation}
g = -tdt^2+a^2(t)(\frac{dr^2}{1-kr^2}+r^2d{\theta}^2+r^2{\sin}^2
{\theta}d{\phi}^2)+\frac{1}{b^2(\tau)}({\tau}d{\tau}^2+dx^2),
\end{equation}
This  metric represents an internal space with $\tau , x$ as coordinates and
$b^{-1}(\tau)$ as the scale factor and a FRW type space-time as $M_4$.
The signature of $ds^2_4$ is determined by the sign of $t$. In other words
$M_4$ has a Euclidean metric whenever $t$ is negative and a Lorentzian
metric when $t$ is positive. So the signature of the metric on $M_4$ changes
whenever $t$ changes sign.
Similarly the metric on $M_2$ has a Euclidean signature for positive
$\tau$ and a Lorentzian one for negative $\tau$ and the signature of the metric
on $M_2$
changes whenever $\tau$ changes sign. We have required $g$ to have only one time-
dimension, that is, we only consider regions where $t<0,\tau<0$ and $t>0,\tau>0$.

Inserting $(8)$ and $(15)$ into Einstein equations $(11)$ and $(12)$ we obtain
the following equations
\begin{equation}
\frac{-3}{t}(\frac{{\dot{a}}^2+kt}{a^2})-\frac{1}{\tau}(b\ddot{b}-{\dot{b}}^2-
\frac{b\dot{b}}{2\tau}) = \frac{1}{2}{\lambda}sign(g_2),
\end{equation}
\begin{equation}
\frac{1}{t}(2\frac{\ddot{a}}{a}+\frac{{\dot{a}}^2+kt}{2at})+\frac{1}{\tau}(
b\ddot{b}-{\dot{b}}^2-\frac{b\dot{b}}{2\tau}) = -\frac{1}{2}{\lambda}sign(g_2),
\end{equation}
\begin{equation}
\frac{3}{t}(\frac{\ddot{a}}{a}+\frac{{\dot{a}}^2+kt}{a^2}-\frac{\dot{a}}{2at})
 = \frac{1}{2}{\lambda}sign(g_2),
\end{equation}
where a dot denotes differentiation with respect to the argument.

Let us restrict ourselves to the case of $k=+1$. Equation (18) contains
$a(t)$ only and hence can be solved easily. A solution is as follows;
for $g_2>0$, (corresponding to $t>0$)
\begin{equation}
a(t)= \frac{1}{H}\cosh(\frac{2}{3}Ht^{3/2}),
\end{equation}
with $12H^2=\lambda$ and therefore (16) and (17) consistently yield
\begin{equation}
\frac{1}{\tau}(b\ddot{b}-{\dot{b}}^2-\frac{b\dot{b}}{2\tau}) = -\frac{3}{4}\lambda.
\end{equation}
A solution to (20) is
\begin{equation}
b(\tau)=\frac{\sqrt{3\lambda}}{2h}\sin(\frac{2}{3}h{\tau}^{3/2}),
\end{equation}
with $h$ being a constant. Thus for positive $t$ (which corresponds to positive
$\tau$) we have a compact (one point compactified) internal space decreasing
in size with time $\tau$, up to
$b^{-1}(\tau=(\frac{3\pi}{4h})^{2/3})=\frac{2h}{\sqrt{3\lambda}}$ while the
four dimensional space has the de Sitter geometry. If we do not deploy the one
point compactification scheme, we will have a bounded non-compact internal
manifold (a two-sphere with its north pole removed). Of course models based on
bounded non-compact internal manifolds have also been proposed, see e.g., \cite{ge}.
For $g_2<0$, (corresponding to $t<0$)
\begin{equation}
a(t)= \frac{1}{H}\cos(\frac{2}{3}H(-t)^{3/2}).
\end{equation}
This solution together with (19) correspond to a scale factor which is
continuous across the signature change hypersurface $t=0$. Now from (16) and
(17) we obtain
\begin{equation}
\frac{1}{\tau}(b\ddot{b}-{\dot{b}}^2-\frac{b\dot{b}}{2\tau}) = \frac{3}{4}\lambda,
\end{equation}
and a solution to this equation is
\begin{equation}
b(\tau) = \frac{\sqrt{3\lambda}}{2l}\sinh{(\frac{2}{3}l(-\tau)^{3/2})},
\end{equation}
in which $l$ is a constant. Thus in this region $M_{4}$ has an
oscillatory scale factor while $M_{2}$ is noncompact and its size
grows with its time $\tau$. The (Euclidean time) $t$-interval
between the big bang and the time at which signature transition
occurred is given by the the first zero of $a(t)$ in (22), that is
$t=(\frac{3\pi}{4H})^{2/3}$.

We can say in conclusion that there are solutions, which accommodate both
compactification of internal dimensions and a signature transition in the metric
of the four dimensional part of the space-time due to the interaction of
electromagnetic fields with gravity. In contrast to previous studies in which a
signature transition is caused by conversion of a space-like dimension to a time-like
one , in our model the transition occurs only by shifting the
time-like dimension. Also in our model there is no dynamical matter field in
the 4-dimensional sector. It should be interesting to study this model in
the framework of quantum cosmology.
\\

I would like to thank H.R. Sepangi , R.W. Tucker and an anonymous referee for
useful comments. Part of this work was done during my visit to the Department
of Physics at Lancaster University.

\end{document}